# A Hitchhiker's Guide through the Bio-image Analysis Software Universe

Robert Haase[1,2], Elnaz Fazeli[3], David Legland[4,5], Michael Doube[6], Siân Culley[7], Ilya Belevich[8], Eija Jokitalo[8], Martin Schorb[9,10], Anna Klemm[11], Christian Tischer[10]

**Affiliations:**

1 DFG Cluster of Excellence "Physics of Life", TU Dresden, Germany

2 Center for Systems Biology Dresden, Germany

3 Biomedicum Imaging Unit, Faculty of Medicine and HiLIFE, University of Helsinki, Finland

4 INRAE, UR BIA, F-44316 Nantes, France

5 INRAE, PROBE research infrastructure, BIBS facility, F-44316 Nantes, France

6 Department of Infectious Diseases and Public Health, City University of Hong Kong, Kowloon, Hong Kong

7 Randall Centre for Cell & Molecular Biophysics, Guy's Campus, King's College London, London SE1 1UL, UK

8 Institute of Biotechnology, Helsinki Institute of Life Science, University of Helsinki, Helsinki, Finland.

9 Electron Microscopy Core Facility, European Molecular Biology Laboratory, Heidelberg, Germany

10 Centre for Bioimage Analysis, European Molecular Biology Laboratory, Heidelberg, Germany

11 VI2 - Department of Information Technology and SciLifeLab BioImage Informatics Facility, Uppsala University, Uppsala, 752 37 Sweden

**Abstract**: Modern research in the life sciences is unthinkable without computational methods for extracting, quantifying and visualizing information derived from biological microscopy imaging data. In the past decade, we observed a dramatic increase in available software packages for these purposes. As it is increasingly difficult to keep track of the number of available image analysis platforms, tool collections, components and emerging technologies, we provide a conservative overview of software we use in daily routine and give insights into emerging new tools. We give guidance on which aspects to consider when choosing the right platform, including aspects such as image data type, skills of the team, infrastructure and community at the institute and availability of time and budget.

**Introduction**

Scientific bio-image analysis software plays a key role in modern life sciences (Levet et al. 2021). New insights are virtually impossible without computational methods for image acquisition, processing, segmentation, feature extraction and visualization. In the past decade, biologists have increasingly applied statistical data analysis of imaging data and machine learning for image processing and particularly for image segmentation, as these allow overcoming the limitations of purely descriptive methods. We also perceive that tools and methods are converging: If a single software platform provides image processing, feature extraction, statistical analysis and visualization, it is superior and preferred to software that is only good in one of those tasks, at least from a user's perspective. The applications are highly diverse and spread across multiple sub-disciplines such as developmental biology, cancer research, immunology, cell and molecular biology, biophysics, agronomy, bioengineering and biomaterials. Often software solutions are created to address a particular analysis challenge in one of those sub-disciplines. As it becomes increasingly hard to keep an overview of existing software, corresponding key applications and targeted scientific questions, we provide a detailed overview of current state-of-the-art software, upcoming next-generation tools and give hints for which aspects to consider when deciding between the many potential software solutions for current bio-image analysis questions.

An early career scientist searching for the right software for their image analysis might have the hardest decision to make. Even if they know search engines specialized for bio-image analysis software such as https://biii.eu or https://bio.tools (BioImage Informatics Index (BIII) 2022; Elixir community 2022), it is hard to make any decision as beginners in the field often don't know the right terms to search for yet. Hence, the glossary provided below may be a good starting point to get an overview of available software and related use-cases. Furthermore, we recommend attending institutional image-analysis courses, e.g. for PhD students in their first year. In addition, getting in touch with senior scientists in their own group, with collaborators and local light or electron microscopy facilities is a good opportunity to find out which software is used in similar projects on campus.

**Glossary**

Inspired by (Adams and Lloyd 1983) we give an overview of the bio-image analysis software universe by means of a glossary of software routinely used by bio-image analysts. To further classify those software related terms, we refer to groups of software as introduced by (Miura and Sladoje 2019): Firstly, image/data analysis algorithms provided in a sustainably reusable fashion are referred to as "components". Secondly, software libraries and standalone applications that combine multiple components are "collections". Thirdly, software that combines multiple components, potentially from multiple collections to solve a given variety of image analysis questions in a standardized form are referred to as "workflow templates". If the software is specific for solving particular scientific questions using given components in one specific assembly, these are called "workflows". We extend this classification with "frameworks" of scientific software which are collections upon which many other software solutions are built. We add "programming languages" that allow assembling components into workflows. We furthermore categorize the presented software in additional categories such as open-source, free of charge and major application categories such as acquisition, registration, segmentation and statistical analysis in Supplementary Table 1. The table also contains properties of the given software such as preferred dimensionality of input image data and typical imaging modality. The software listed in the following glossary were selected to reflect long-term available, reliable, sustainably maintained and supported software solutions. We analysts often have to have a conservative perspective on existing software as we need to rely on established, reliable and maintained software to build workflows for our collaborators and trust the given software to be still available in 5-10 years allowing reproducible image data analysis. To this end, the number of citations served as a criterion to select software but we also considered software packages that have been available for about 5-10 years with continuous maintenance and reliable support by a vivid community. A less formal criterion we applied for selecting software was considering which tools a starting bio-image analyst would have benefitted from being aware of in order to follow a conversation at one of the Network of European Bio-Image Analysts (NEUBIAS) meetings. The given description of the glossary items highlights the main application of the software and its relationships with other glossary items. While the glossary focuses on general software and terms used in the field, this should not hide the large number of software or plugins developed for a specific task or context. In the domain of plant science, the quantitative plant initiative (Lobet, Draye, and Périlleux 2013) proposes a curated list of software solutions that may also be useful to know. Also for lightsheet microscopy, there is a specific list of software for acquisition and analysis available (Gibbs et al. 2021).There are a large number of software tools and applications that have been specifically developed for the cryo-electron microscopy (cryoEM); a comprehensive list of software for the cryoEM community can be found from wikibooks ("Software Tools For Molecular Microscopy" 2006).

**3D ImageJ Suite** (Ollion et al. 2013) is a collection of ImageJ plugins for filtering, segmentation and analysis of geometry, shape, and spatial organization of objects in 3D images.

**3D Slicer** (Fedorov et al. 2012) is an image processing software based on the ITK library focused on medical imaging with 3D surface extraction, rendering and analysis capabilities. It is increasingly used for visualizing and analyzing 3D structures such as cells, tissues and organs in microscopy data.

**ANTs** (Avants et al. 2011), or Advanced Normalization Tools, is a collection of methods for image registration, segmentation, and analysis, mostly developed in the context of neuro-imaging and the comparison of cohorts. ANTs depends on ITK, an image processing library to which ANTs developers contribute.

**Arivis Vision 4D** (arivis AG, Rostock, Germany) is an image analysis software for processing multi-channel 2D, 3D and 4D data, focused but not limited to microscopy data. It is scalable, supports processing big image data, and has intuitive image stitching and alignment tools.

**Amira-Avizo** (Thermo Fisher Scientific Inc, Waltham, MA, United States) is a 2D-5D image processing, visualization and analysis software. It can be customized using Python and MATLAB and offers additions for incorporating artificial intelligence.

**BigDataViewer** (Pietzsch et al. 2015) is an *n*-dimensional image viewer component for slicing volumes in arbitrary directions. The Fiji plugin can handle terabyte-sized image data composed of multiple channels and time points.

**BigStitcher** (Hörl et al. 2019) is an automated and interactive image registration / fusion Fiji plugin capable of handling terabyte sized image data. It is based on the BigDataViewer.

**BioFormats** (Linkert et al. 2010) is an image file format interoperability library which serves multiple image analysis software applications such as Fiji, Omero and QuPath to load image data from many formats and vendors.

**Blender** (Blender Foundation 2022) is a 3D surface rendering, modeling and visualization software with Python scripting, simulation, and video editing capabilities. The home of Blender is in design and arts, and it is increasingly used for microscopy image data visualization.

**BoneJ** (Domander, Felder, and Doube 2021) is a collection of image processing operations and ImageJ plugins for skeletal / bone image analysis. It is used often in the soil, food and materials science communities. Some of the tools were updated to work with ImageJ2.

**C/C++** are programming languages traditionally used in computing. Most operating systems are programmed in C and C++. Furthermore, many Python and also some Java libraries contain components and collections of [image] processing routines written in these languages because C and C++ offer higher performance.

**CATMAID** (Saalfeld et al. 2009) is a web application to navigate, share and collaboratively annotate massive volume image data sets.

**CCP-EM** (Burnley, Palmer, and Winn 2017) the Collaborative Computational Project for electron cryo-microscopy is a community guiding the users of cryo-EM software tools as well as developers of software packages and file formats.

**CCPi** ("CCPi Tomographic Imaging" 2022) the Collaborative Computational Project in Tomographic Imaging provides a collection of software tools for tomographic imaging and reconstruction.

**CellPose** (Stringer et al. 2021) is a deep-learning based segmentation algorithm for biological structures such as cell and cell nuclei in microscopy images. It is accessible as a Python library. Third parties maintain CellPose plugins for QuPath and Fiji.

**CellProfiler** (McQuin et al. 2018) is an image analysis software application with graphical user interface (GUI) for user-friendly configuration of standardized image analysis workflows focusing on high-throughput microscopy imaging data of cells with capabilities for extracting tabular image feature data in high-performance-computing environments.

**CellProfiler Analyst** (Jones et al. 2008) is a data exploration software for further visualization and analysis of tabular data produced with CellProfiler. It offers advanced plotting, dimensionality reduction and machine learning based object classification for dealing with big data as it is common in pharmaceutical research.

**DeconvolutionLab2** (Sage et al. 2017) is a collection of image deconvolution algorithms accessible as standalone command-line interface and as user-friendly ImageJ plugin.

**Dragonfly** (ORS, Montréal, Canada) is a powerful standalone software featuring an extensive set of tools for image processing, segmentation and 3D visualization.

**Drishti** (Limaye 2012; Hu, Limaye, and Lu 2020) is a visualization tool for 3D pixel data, which has been extended with segmentation and measurement tools.

**Elastix** (Klein et al. 2010) is a standalone command-line tool for registration of 2D and 3D image data based on the ITK library. A Python compatible interface, SimpleElastix, is available as well.

**EMAN2** (Tang et al. 2007) is a software application focusing on CryoEM, covering techniques such as single particle analysis, cryo-electron tomography or sub-tomogram averaging.

**fairSIM** (Müller et al. 2016) is an ImageJ plugin for reconstructing structured illumination microscopy super-resolution images from raw data.

**Fiji** (Schindelin et al. 2012) is an image analysis software based on ImageJ and a collection of ImageJ- and ImageJ2 compatible plugins focusing on general-purpose image analysis in the life sciences. It Is scriptable using multiple programming languages compatible with the Java ecosystem, extensible and capable of handling big image data through integration of components such as ImgLib2 and BigDataViewer.

**Groovy** ("The Apache Groovy Programming Language" 2022) is a scripting language that can be used for automating image analysis routines in QuPath and Fiji.

**Gwyddion** (Nečas and Klapetek, 2012) is a modular program for scanning probe microscopy (SPM) data visualization and analysis, primarily focused on the analysis of altitude maps such as obtained by atomic force microscopes.

**Huygens** (Scientific Volume Imaging B.V., Hilversum, Netherlands) is an image processing software dedicated to deconvolution of 3D stacks from fluorescence microscopy, potentially multi-channel and time-lapse data.

**Icy** (de Chaumont et al. 2012) is an image analysis software focusing on general purpose image-analysis in the life sciences compatible with ImageJ. Icy is scriptable using JavaScript and a visual programming approach using so called protocols.

**Ilastik** (Berg et al. 2019) is an image analysis software offering easy-to-use machine learning capabilities for image segmentation, object classification, object tracking and statistical analysis of microscopy image data. Ilastik classifiers can be used from Fiji and CellProfiler. Furthermore, it supports execution on high-performance-computing clusters.

**ImageJ** (Schneider, Rasband, and Eliceiri 2012) is an image analysis software and framework for image analysis algorithms integrated in Fiji, Icy, MicroManager, QuPath and others. We conservatively estimate tens or hundreds of thousands of plugins and scripts have been developed in its 20+ year history making it one of the most important platforms for image analysis in the life sciences.

**ImageJ2** (Rueden et al. 2017) is a modern rewrite of the ImageJ codebase with focus in scientific image processing and analysis of big image data. It serves as an extensible platform underlying Fiji and other software platforms in the life sciences.

**ImageJ Macro** ("Macro Language" 2022) is a limited programming language specific to the ImageJ platform useful for automating image processing routines.

**Image.sc** (Rueden et al. 2019) is an online discussion forum based on the Discourse platform ("Discourse - Civilized Discussion" 2022) that serves as questions & answers forum for many open source projects from the image science field. It plays a key role in knowledge exchange and community support for many open source bio-image analysis software projects. See Figure 1 for a list of community partners.

**Imaris** (Oxford Instruments, Oxon, United Kingdom) is an image processing and visualization software supporting 3D volume rendering and quantitative analysis. Through extra modules it is interoperable with Fiji, Python and MATLAB.

**ImgLib2** (Pietzsch et al. 2012) is an image processing framework and collection of algorithms. It is the basis for software such as BigDataViewer, BigStitcher, ImageJ2, Fiji, Knime and others to handle terabyte-sized big image data.

**IMOD** (Kremer, Mastronarde, and McIntosh 1996) is an image processing, modeling and visualization software collection for electron microscopy. Aside from command line tools for image processing, it offers a GUI for reconstruction, registration and segmentation of data.

**ITK** (Yoo et al. 2002) is an image registration and segmentation algorithm collection and library with a long history in medical imaging. It is the underlying framework for tools such as 3D Slicer, Elastix, ANTs, and ITK Snap.

**ITK Snap** (Yushkevich et al. 2006) is a software application specifically for segmentation and surface rendering of 3D medical imaging datasets based on ITK.

**Java** is the programming language Icy, ImageJ, Fiji, QuPath and compatible plugins are written in. It is also interoperable with Imaris and MATLAB.

**JavaScript** is a scripting language used for automation of image analysis routines in Icy, ImageJ and Fiji. It is also the most popular web programming language world wide ("Stack Overflow Developer Survey 2021" 2021).

**Jupyter Notebooks** (Kluyver et al. 2016) is an interactive, cloud compatible programming environment suitable for image data analysis, statistics and scientific plotting. It is a key component for reproducible data science in the scientific Python ecosystem and is extensively used for documentation and training.

**Jython** is a Java-compatible scripting language based on the syntax of Python 2. It can be used for automation of image analysis routines in Fiji but is technically *not* compatible with numpy, scipy, scikit-image and other Python-based libraries. It is compatible with Java-based components.

**KNIME** (Berthold et al. 2008) is a visual and interactive programming environment focusing on data science with image analysis and machine learning capabilities. Its image processing capabilities are based on ImageJ, ImageJ2, SciJava and Imglib2.

**Knossos** (Helmstaedter, Briggman, and Denk 2011) is an image visualization and annotation software for large connectomics (electron microscopy) data extensible using Python modules.

**Leica Application Suite X** (Leica microsystems GmbH, Wetzlar, Germany) is a software for microscope control, image acquisition, visualization and analysis. It offers modules for computational clearing and deconvolution (lightning), Fluorescence lifetime, FRET, and FCS analysis, CARS calculations, 2D and 3D measurements.

**MATLAB** (Mathworks, Natick, MA, United States) is a software environment for numeric computing that provides a multi-paradigm programming language and a number of dedicated applications and toolboxes, e.g. for image processing, computer vision, statistics and machine learning. It can be extended using Java libraries.

**Matplotlib** (Hunter 2007) is a scientific plotting and image visualization collection commonly used for image data science by the Python community.

**MicroManager** (Edelstein et al. 2010) is a microscope control software with built-in image processing capabilities based on ImageJ. It can be scripted using the BeanShell language and recently using Python (Pinkard et al. 2021).

**Microscopy Image Browser** (Belevich et al. 2016) is a MATLAB-based software for advanced image processing, segmentation, quantification, and visualization of multi-dimensional light and electron microscopy datasets. It works with BioFormats, allows batch processing operations and can be directly linked to Fiji.

**MorphographX** (Barbier de Reuille et al. 2015) is a software for visualization and analysis of 4D datasets. It focuses on the analysis of organ growth from 4D live-imaging confocal data of plants. Various algorithms implemented in MorphographX extract surfaces from 3D data and post-process the intensities along those surfaces, which can be seen as an efficient 2.5 dimensional approximation of 3D quantification.

**MorphoLibJ** (Legland, Arganda-Carreras, and Andrey 2016) is a collection of methods and plugins for ImageJ implementing mathematical morphology operations such as dilation, opening, watershed and reconstruction as well as methods for quantitative analysis of label images.

**NanoJ** (Laine et al. 2019) is a toolbox of ImageJ plugins for super-resolution microscopy processing and analysis tasks, including drift correction and channel registration. It also incorporates the widely-used SRRF method for live-cell super-resolution image reconstruction (Gustafsson et al. 2016).

**NeuronJ** (Meijering et al. 2004) is an ImageJ plugin for neurite tracing and analysis.

**NIS-Elements** (Nikon, Tokyo, Japan) is a software for microscope control, computer-assisted image acquisition and analysis. It integrates artificial intelligence solutions for de-blurring, segmentation and image restoration. Image analysis components can be combined to a workflow within a visual programming environment.

**Numpy** (Harris et al. 2020) is a Python library and a collection of efficient array processing algorithms. It is among the most used Python libraries in the world ("Python Package Index Download Statistics" 2021) and the basis for many image processing components and collections in the Python ecosystem.

**ParaView** (Ahrens, Geveci, and Law 2005) is a software for vector and surface data analysis and visualization based on the ITK library.

**Python** is a programming language, potentially the most popular language in science and surely among the top used programming languages in general ("Stack Overflow Developer Survey 2020" 2020). It is commonly used to assemble various image processing, data analysis and visualization libraries in scientific workflow.

**OrientationJ** (Püspöki et al. 2016) is an ImageJ plugin to characterize the orientation and isotropy properties of regions of interest in images.

**Omero** (Allan et al. 2012) is a research data management solution for microscopy image data. It was initially developed to facilitate analysis of large amounts of high-throughput imaging data. Omero can be used as a remote-server storing image data that is highly interoperable with other software such as CellProfiler, Fiji and QuPath.

**OpenCV** (Bradski 2000) is a collection of image analysis components that includes several hundred computer vision algorithms. OpenCV focuses on 2D+time imaging data acquired with video cameras and has also many applications in microscopy.

**QuPath** (Bankhead et al. 2017) is an image analysis software for quantitative pathology. It allows visualization and analysis of large 2D slide scanner imaging data of histological slices. Its user-friendly GUI offers tools for manual annotation, machine-learning based tissue classification and deep-learning based cell segmentation. It is extensible using Java-based plugins and scriptable using the Groovy programming language. It is interoperable with Omero and BioFormats.

**R** ("The R Project for Statistical Computing" 2022) is a programming language for statistical computing and plotting. It is commonly used for the downstream statistical analysis of the output of image analysis packages. R-packages also exist for image processing (Pau et al. 2010).

**RELION** (Scheres 2012), or REgularised LIkelihood OptimisatioN, is a software package for cryo-EM structure determination processing data from single particle or tomography experiments.

**RStudio** ("The R Project for Statistical Computing" n.d.) is a standalone application allowing interactive programming using the R language. Users can view existing variables, manipulate tables and plots.

**SciJava** ("SciJava" 2022) is a collection of image analysis data structures and algorithms such as ImgLib2 and serves as the basis for ImageJ2.

**Scikit-image** (van der Walt et al. 2014) is a general purpose collection of scientific image analysis algorithms based on numpy and scipy. Image analysis workflows using scikit-image can be written in Python and it is commonly used with jupyter notebooks.

**Scikit-learn** (Pedregosa et al. 2011) is a collection of Python-based algorithms for machine learning commonly used in the context of image for pixel, object and image classification.

**SCILS** (Bruker, Billerica, MA, United states) is a software for analysis of mass-spectrometry imaging (MSI) data, including machine learning algorithms and tools for visualizing ion images and mass spectra.

**Scipy** (van der Walt et al. 2014; Virtanen et al. 2020) is a collection of algorithms for scientific data processing, simulation, optimization and analysis. It serves as the basis for other software such as scikit-image.

**SerialEM** (Mastronarde 2005) is an acquisition software for a variety of transmission electron microscopes. It provides different means of automation through navigation, a built-in scripting language and Python integration. Typical applications are electron tomography, large areas for 3-D volume imaging from serial sections or single-particle cryo-EM.

**Single Neurite Tracer** (Arshadi et al. 2021) is a Fiji plugin for processing three-dimensional, multi-channel, timelapse data to trace neurites including analysis and plotting.

**SMAP** (Ries 2020) is a MATLAB-based framework for 2D and 3D single-molecule localization microscopy analysis encompassing tasks such as molecule localization, image rendering, and quantitative analysis.

**SR-Tessler** (Levet et al. 2015) is a standalone software for quantitative analysis of localization-based super-resolution microscopy data.

**StackReg** (Thévenaz, Ruttimann, and Unser 1998) is an ImageJ plugin for 2D+time image registration. It is also commonly used for other types of image registration, e.g. for alignment of slices in 3D image stacks.

**StarDist** (Schmidt et al. 2018; Weigert et al. 2020) is a deep-learning based Python library for segmenting star-shaped objects such as cell nuclei which is also available as plugins for Fiji and QuPath.

**ThunderSTORM** (Ovesný et al. 2014) is an ImageJ plugin for automated processing, analysis and visualization of data acquired by single-molecule localization microscopy. ThunderSTORM is at the moment of writing not actively maintained and may in the future be replaced by other solutions.

**TomoPy** (Gürsoy et al. 2014) is an open-source Python package for processing tomography data and image reconstruction. It is mainly used for X-Ray tomography.

**Tomviz** (Levin et al. 2018) is a software package tailored for processing, visualization, and analysis of 3D tomographic data acquired with transmission electron microscopy. It is compatible with Python scripting to accommodate custom algorithms.

**TrackMate** (Tinevez et al. 2017) is a Fiji plugin for object tracking in 2D+t and 3D+t image data. It comes with advanced plotting, track visualization and cell lineage tree visualization tools. It is extensible using Java-based plugins and scriptable using advanced scripting languages in Fiji such as Groovy, JavaScript and Jython.

**Trainable WEKA Segmentation** (Arganda-Carreras et al. 2017) is a user-friendly ImageJ plugin for pixel classification using various machine learning techniques based on the Waikato Environment for Knowledge Analysis (Witten et al. 2016).

**TrakEM2** (Arganda-Carreras et al. 2017; Cardona et al. 2012) is a Fiji plugin for registration, stitching and management of large-scale electron microscopy data which offers tools for segmentation and reconstruction of objects such as neurons in 3D.

**Zen** (Zeiss AG, Oberkochen, Germany) is a software for microscope control, image acquisition, visualization and analysis. It offers components for image stitching, registration and segmentation.

### Emerging software

Apart from our conservative view on the field, we also perceive recent software developments which presumably will become part of the above glossary within the next 5-10 years. Most prominently, deep learning approaches are flooding our field with promising image processing components especially for image restoration (Weigert et al. 2018; Krull, Buchholz, and Jug 2018; Batson and Royer 2019) and cell / nuclei segmentation (Schmidt et al. 2018; Weigert et al. 2020; Aigouy et al. 2020; Stringer et al. 2021), classification and tracking (Mathis et al. 2018) within complex scenes. Readers interested in an extended list of new application and ready-to-use deep-learning models are referred to (Belthangady and Royer 2019) and BioImage. IO 2022.

Those deep-learning based components rely on technical frameworks such as TensorFlow (Tensorflow-Developers 2021) and PyTorch (Paszke et al. 2019) which are not directly accessible to end-users. Multiple user-friendly GUIs were recently developed offering modern deep-learning tools to a wide target audience (Gómez-de-Mariscal et al. 2021; von Chamier et al. 2021; Ouyang et al. 2019; Belevich and Jokitalo 2021). User-friendly deep-learning based image processing is also already integrated within some of the applications listed in the glossary, namely <u>ilastik, Microscopy Image Browser</u> and almost all of the commercial software packages.

In the same context, the napari project (Sofroniew et al. 2021) is bridging the Python community towards the life scientists community by offering automatically generated, user-friendly GUIs to the most recent deep-learning and data science components and strives to become a major framework of the bio-image analysis community. From a Python community perspective, napari is already a game changer as it brings widely usable *n*-dimensional viewing to the otherwise scripting centered Python community (Perkel 2021). From a wider perspective, more image visualization tools have been published recently and show high potential to become major players within the next decade since, compared to current default solutions, they provide opportunities to processing big image data and applying deep learning to microscopy image data (Vergara et al. 2021; Chiaruttini et al. 2022).

Processing big image data, in the form of large 3D image stacks, long 3D+time data or large collections of 2D or 3D image data sets, is also a hot topic where new tools developed using remote-data, remote-computing and network-based approaches are emerging (Afgan et al. 2018; Marée et al. 2016; Rubens et al. 2020; Tischer et al. 2021; Rubens et al. 2019; Boergens et al. 2017; Rocklin 2015) and also semi-commercial solutions are appearing such as Apeer (Zeiss AG, Oberkochen Germany). Graphics-Processing-Unit (GPU)-accelerated classical image processing (Ryosuke Okuta, Yuya Unno, Daisuke Nishino, Shohei Hido, Crissman Loomis 2017; Haase, Royer, et al. 2020; Dzmitry Malyshau, Kai Ninomiya, Brandon Jones (Editors) 2022; Haase, Jain, et al. 2020) will play a major role for overcoming current limitations concerning processing times for large image data. From our perspective, such big-data capable solutions will also facilitate analyzing spatial relationships in biological specimen and tissues using modern data-science approaches in the context of spatial-omics and transcriptomics (Stoltzfus et al. 2020; Palla et al. 2021; Axelrod et al. 2021; Haase 2021; Solorzano, Partel, and Wählby 2020; Pielawski et al. 2022).

For single molecule localization microscopy analysis, improved methods for molecule detection and localization are in active development (Sage et al. 2019). In the field of super-resolution microscopy more broadly, there is also a focus on developing user-friendly methods for ensuring the fidelity of reconstructed images (Culley et al. 2018; Ball et al. 2015; Marsh et al. 2021).

Last but not least, new file formats and solutions for research image data management (Miles et al. 2020; Moore et al. 2021; Stephan Saalfeld, Igor Pisarev, Philipp Hanslovsky, John Bogovic, Andrew Champion, Curtis Rueden, John Kirkham 2017) are under development and we expect those to have a huge impact on how analysts handle image data within the next decade.

**Aspects to consider when choosing bio-image analysis software for your research**

The choice of the right bio-image analysis software is closely related to the purpose of a given research project and more broadly to the field a research group is working in. We suggest becoming confident in a single software that broadly fits the planned research needs instead of switching the used software from project to project just because a single feature may be more accessible or more accurate in another software. Getting to know software and maintaining expertise comes at a high cost when numerous potentially incompatible platforms are used.

Interoperability between software is another key feature to consider. For example, we discourage using software that comes with proprietary custom file formats and suggest using broadly available file formats instead. Most prominently, there are some software packages with custom formats for project files. We recommend making sure these project files can be accessed with common text editors and contain human-readable text that is based on standard formats such as XML, JSON, YAML or CSV. It also appears beneficial if software has capabilities for workflow automation or, in the case of plugins, can be automated using the platform they can be integrated with. For example, software with a great built-in segmentation algorithm can become a major bottleneck if the algorithm cannot be integrated with other software, e.g. for pre-processing, post-processing, feature-extraction and statistical analysis.

Striving for reproducible bio-image analysis workflows with minimal manual interaction steps is key for analyzing large amounts of image data leading to insights cemented by appropriate measurements exhibiting statistical power. If the software supports forming and properly documenting such automated workflows, reproducibility and interpretability of results can be ensured (Miura and Nørrelykke 2021).

Other technical aspects such as big data capabilities play a key role, especially when new microscopy techniques potentially producing more and more data are published every year. Many software packages claim the ability to work with big data, but often refer to visualization only or refer to big data as many images with a size of megabytes to gigabytes each. On the other hand, software packages capable of processing big volumetric image data in the range of terabytes and petabytes to produce quantitative analysis results are still rare and often limited in other aspects, for example image data dimensionality or imaging modality. We see more and more web-based solutions being published diminishing the need to buy expensive computational hardware, to train and execute neural network architectures. When using web-based solutions in the cloud, institutional, national and international laws have to be respected. Additional technical burdens hinder the wide adoption of cloud computing at the moment. For example, uploading multiple terabytes of imaging data from a European institute to an American computing server is not just challenging from a legal but also from the file transfer bandwidth perspective. We assume these burdens will fall in the next decade and the technology will become available to more and more researchers as the benefits of using it outweigh the risks. Hence, choosing a software that is interoperable with cloud-computing and cloud-storage technologies appears a future-proof approach.

Many image analysis tasks have a substantial number of solutions that have been developed, and it can often be unclear which solution is most appropriate for a user's specific problem. Several image analysis fields have established benchmarking challenges, whereby software is applied to exemplar datasets and performance is automatically and independently assessed. Such challenges exist for cell tracking (Ulman et al. 2017; "Cell Tracking Benchmark" 2022), cell segmentation (Arganda-Carreras et al. 2015), electron microscopy image segmentation (Wei et al. 2020) and single molecule localization (Sage et al. 2019). These provide users with a quantitative comparison of the state-of-the-art, along with test datasets and guidance for quality reporting.

Community aspects should also be taken into account when choosing the right image analysis software. A key role in bio-image analysis for microscopy is played by the Image Science community https://image.sc, an online forum where developers and users of most software listed in the glossary are actively supporting each other by providing support and feedback (Figure 1). Before using a software mid-/long-term, users may want to explore this forum and other online platforms to figure out how actively supported the software is by a broader community. Furthermore, some software communities hold regular virtual community meetings, where users and plugin-developers can get in touch with core-developers to exchange ideas, use-cases and receive support. The weekly community meeting of the napari community and the open office hours of the CellProfiler community shall be highlighted here as well-appreciated examples. For staying up-to-date with new developments in bio-image analysis software, following new media channels such as the NEUBIAS Academy YouTube channel (NEUBIAS 2020) and the @Talk_BioImg Twitter bot which retweets posts containing the #BioImageAnalysis hashtag should be considered as well for an audience with general interest in the field, e.g. postdocs and core-facility staff working on applied bio-image analysis and image data science. In addition, the Global Bioimaging infrastructure is organizing image analysis courses and providing a training resource for core facility staff and image analysis community.

From a group leader's and an institutional decision maker's perspective, guiding scientists towards using a common software platform makes sense. The more local collaborators work with the same software, and maybe just use different plugins, the more they can support each other and exchange knowledge. If it is apparent that a majority of the group or institute members are using the same software, an institute can strengthen this community by inviting the developers of that software annually for courses and seminars. Building this bridge between users and developers is of mutual benefit: While the users receive support and training from experts, the experts can establish collaborations with power users leading to scientific publications. These applied-science publications are key to grant applications and sustainable maintenance of research software.

**Conclusions and Perspectives**

Comparing the bio-image analysis software universe one decade ago to its state now clearly shows that it is expanding. Multiple huge ecosystems grew from a small number of general-purpose software developed more than a decade ago. Recently, many of those platforms are increasing their efforts to build bridges, which is appreciated from our bio-image analysis workflow

designers` perspective. Users are well-advised to use interoperable, established, sustainably maintained software packages for their research. Also keeping an eye on recent developments, especially in the deep learning context, makes a lot of sense, while staying a brave scientist and questioning those new methods. Such critical thinking is necessary to solidify knowledge about new methods before eventually adopting them as community standards. Last but not least we want to emphasize that the decision on which software to use for specific research projects should be made in groups. Getting in touch with local, regional and global experts and discussing advantages and disadvantages is the right path in a growing universe of bio-image analysis software.

**Acknowledgements**

We would like to thank Mafalda Sousa (I3S, Porto), Ignacio Arganda-Carreras (Universidad del Pais Vasco, Donostia-San Sebastian), Martin Jones (The Francis Crick Institute, London) for reviewing the manuscript and providing constructive feedback prior to submission.

RH acknowledges support by the Deutsche Forschungsgemeinschaft under Germany's Excellence Strategy - EXC2068 - Cluster of Excellence Physics of Life of TU Dresden. EF, IB and EJ acknowledge support by Biomedicum Imaging Unit and Electron Microscopy Unit, Helsinki University, as a part of Biocenter Finland infrastructure. CT acknowledges support by grant number 2020- 225265 from the Chan Zuckerberg Initiative DAF, an advised fund of Silicon Valley Community Foundation. SC acknowledges support from a Royal Society University Research Fellowship. AK was funded via the BioImage Informatics Facility, a unit of the National Bioinformatics Infrastructure Sweden NBIS, with funding from SciLifeLab, National Microscopy Infrastructure NMI (VR-RFI 2019-00217), and the Chan-Zuckerberg Initiative.

# Figures

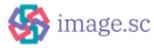

Figure 1: Screenshot of the image.sc forum in April 2022 showing the logos of the community partners and related communities. Listed open-source projects provide online support for their software on this platform, which might be a key-criterion when deciding for which software to use.

**Tables**

| Software | Standalone Application with GUI | Command Line interface | Online platform | Plugin | Framework | Extensible | Scriptable | Programming language | Component | Collection | Workflow | Open Source | Free of charge | Acquisition | Reconstruction | Deconvolution | Denoising | Background removal | Registration | Stitching | Manual annotation | Segmentation | Surface extraction | Object Tracking / cell lineage tracing | Quantification / feature extraction | Statistical analysis | Visualization | Plotting | Preferred data dimensionality | Preferred modality |
|---|---|---|---|---|---|---|---|---|---|---|---|---|---|---|---|---|---|---|---|---|---|---|---|---|---|---|---|---|---|---|
| 3D ImageJ Suite | | | | x | | | x | | | | | | | | | | | | | | | x | | | x | | x | | 3D+c | |
| 3D Slicer | x | | | x | x | x | | | | | | | | | | | | | | | | | | x | x | | x | | 3D | MRI/CT/PET |
| ANTs | | | | | | | | | | | | | | | | | | | | | | | | x | x | | 3D | MRI |
| Arivis Vision 4D | x | | | x | | x | | | | | | | | | | | | | | | x | x | x | x | x | x | 3D+t+c | |
| Amira-Avizo | x | | | x | | x | | | | | | | | | | | | | | | x | x | x | x | x | x | 2D-3D+t+c | |
| BigDataViewer | | | x | | | x | | | x | | | | | | | | | | | | | | | | x | | 3D+t+c | |
| BigStitcher | | | x | | | x | | | x | | | | | | | | | | | | | | | | x | | 3D+t+c | |
| BioFormats | | | x | | | x | | | x | | | | | | | | | | | | | | | | | | nD | |
| Blender | x | x | | x | | x | | | | | | | | | | | | | | | x | | | x | | | 3D+t | |
| BoneJ | | | x | x | x | | | | | | | | | | | | | | | | x | | | x | | | 3D | XMT, CT |
| C/C++ | | | | | | | x | | | | | | | | | | | | | | | | | | | | | |
| CATMAID | | x | | | | | | | | | | | | | | | | | | | | | | | | | | |
| CCP-EM | x | x | | x | x | | | | | | | | | | | | | | | | | | | | x | | | cryo-EM |

Supplementary Table 1: List of the software terms from the glossary with categorical information and interactive filtering capabilities available for download https://cloudstore.zih.tu-dresden.de/index.php/s/AaLoe38QJN5aN48 . The software is classified according to a user interface in Standalone Application with GUI, Command Line interface and Online platforms. Furthermore, plugins that are parts of other software are differentiated from frameworks which serve as basis for others and software that is extensible using plugins. Scriptable software and programming languages are specified as well. A classification in Components, Collections and Workflow as suggested by (Miura and Sladoje 2019) is also provided. We differentiate between Open Source software and software that is available free of charge. Image processing features are categorized according to the categories Acquisition, Reconstruction, Deconvolution, Denoising, Background removal, Registration, Stitching, Manual annotation, Segmentation, Surface extraction, Object Tracking / cell lineage tracing, Quantification / feature extraction, Statistical analysis, Visualization and Plotting. The given preferred data dimensionality and preferred modality specify which use-cases a software was made for. The software may be applicable to data of other dimensionality and from other modalities.

# References

Adams, Douglas, and John Lloyd. 1983. *The Meaning of Liff*. Pan.

Afgan, Enis, Dannon Baker, Bérénice Batut, Marius van den Beek, Dave Bouvier, Martin Cech, John Chilton, et al. 2018. "The Galaxy Platform for Accessible, Reproducible and Collaborative Biomedical Analyses: 2018 Update." *Nucleic Acids Research* 46 (W1): W537–44.

Ahrens, James, Berk Geveci, and Charles Law. 2005. "ParaView: An End-User Tool for Large-Data Visualization." In *Visualization Handbook*, 717–31. Elsevier.

Aigouy, Benoit, Claudio Cortes, Shanda Liu, and Benjamin Prud'Homme. 2020. "EPySeg: A Coding-Free Solution for Automated Segmentation of Epithelia Using Deep Learning." *Development* 147 (24). https://doi.org/10.1242/dev.194589.

Allan, Chris, Jean-Marie Burel, Josh Moore, Colin Blackburn, Melissa Linkert, Scott Loynton, Donald Macdonald, et al. 2012. "OMERO: Flexible, Model-Driven Data Management for Experimental Biology." *Nature Methods* 9 (3): 245–53.

Arganda-Carreras, Ignacio, Verena Kaynig, Curtis Rueden, Kevin W. Eliceiri, Johannes Schindelin, Albert Cardona, and H. Sebastian Seung. 2017. "Trainable Weka Segmentation: A Machine Learning Tool for Microscopy Pixel Classification." *Bioinformatics* 33 (15): 2424–26.

Arganda-Carreras, Ignacio, Srinivas C. Turaga, Daniel R. Berger, Dan Cireşan, Alessandro Giusti, Luca M. Gambardella, Jürgen Schmidhuber, et al. 2015. "Crowdsourcing the Creation of Image Segmentation Algorithms for Connectomics." *Frontiers in Neuroanatomy* 9 (November): 142.

Arshadi, Cameron, Ulrik Günther, Mark Eddison, Kyle I. S. Harrington, and Tiago A. Ferreira. 2021. "SNT: A Unifying Toolbox for Quantification of Neuronal Anatomy." *Nature Methods* 18 (4): 374–77.

Avants, Brian B., Nicholas J. Tustison, Gang Song, Philip A. Cook, Arno Klein, and James C. Gee. 2011. "A Reproducible Evaluation of ANTs Similarity Metric Performance in Brain Image Registration." *NeuroImage* 54 (3): 2033–44.

Axelrod, Shannon, Matthew Cai, Ambrose Carr, Jeremy Freeman, Deep Ganguli, Justin Kiggins, Brian Long, Tony Tung, and Kevin Yamauchi. 2021. "Starfish: Scalable Pipelines for Image-Based Transcriptomics." *Journal of Open Source Software* 6 (61): 2440.

Ball, Graeme, Justin Demmerle, Rainer Kaufmann, Ilan Davis, Ian M. Dobbie, and Lothar Schermelleh. 2015. "SIMcheck: A Toolbox for Successful Super-Resolution Structured Illumination Microscopy." *Scientific Reports* 5 (November): 15915.

Bankhead, Peter, Maurice B. Loughrey, José A. Fernández, Yvonne Dombrowski, Darragh G. McArt, Philip D. Dunne, Stephen McQuaid, et al. 2017. "QuPath: Open Source Software for Digital Pathology Image Analysis." *Scientific Reports* 7 (1): 16878.

Barbier de Reuille, Pierre, Anne-Lise Routier-Kierzkowska, Daniel Kierzkowski, George W. Bassel, Thierry Schüpbach, Gerardo Tauriello, Namrata Bajpai, et al. 2015. "MorphoGraphX: A Platform for Quantifying Morphogenesis in 4D." *eLife* 4 (May): 05864.

Batson, Joshua, and Loic Royer. 2019. "Noise2Self: Blind Denoising by Self-Supervision." http://arxiv.org/abs/1901.11365.

Belevich, Ilya, Merja Joensuu, Darshan Kumar, Helena Vihinen, and Eija Jokitalo. 2016. "Microscopy Image Browser: A Platform for Segmentation and Analysis of Multidimensional Datasets." *PLoS Biology* 14 (1): e1002340.

Belevich, Ilya, and Eija Jokitalo. 2021. "DeepMIB: User-Friendly and Open-Source Software for Training of Deep Learning Network for Biological Image Segmentation." *PLoS Computational Biology* 17 (3): e1008374.

Berg, Stuart, Dominik Kutra, Thorben Kroeger, Christoph N. Straehle, Bernhard X. Kausler,

Carsten Haubold, Martin Schiegg, et al. 2019. "Ilastik: Interactive Machine Learning for (bio)image Analysis." *Nature Methods* 16 (12): 1226–32.

Berthold, Michael R., Nicolas Cebron, Fabian Dill, Thomas R. Gabriel, Tobias Kötter, Thorsten Meinl, Peter Ohl, Christoph Sieb, Kilian Thiel, and Bernd Wiswedel. 2008. "KNIME: The Konstanz Information Miner." In *Data Analysis, Machine Learning and Applications*, 319–26. Studies in Classification, Data Analysis, and Knowledge Organization. Berlin, Heidelberg: Springer Berlin Heidelberg.

"BioImage Informatics Index." 2022. 2022. https://biii.eu.

Blender Foundation. 2022. "Blender.org - Home of the Blender Project - Free and Open 3D Creation Software." 2022. https://www.blender.org/.

Boergens, Kevin M., Manuel Berning, Tom Bocklisch, Dominic Bräunlein, Florian Drawitsch, Johannes Frohnhofen, Tom Herold, et al. 2017. "webKnossos: Efficient Online 3D Data Annotation for Connectomics." *Nature Methods* 14 (7): 691–94.

Bradski, G. 2000. "The OpenCV Library." In *Dr Dobb's Journal of Software Tools*.

Burnley, Tom, Colin M. Palmer, and Martyn Winn. 2017. "Recent Developments in the CCP-EM Software Suite." *Acta Crystallographica. Section D, Structural Biology* 73 (Pt 6): 469–77.

Cardona, Albert, Stephan Saalfeld, Johannes Schindelin, Ignacio Arganda-Carreras, Stephan Preibisch, Mark Longair, Pavel Tomancak, Volker Hartenstein, and Rodney J. Douglas. 2012. "TrakEM2 Software for Neural Circuit Reconstruction." *PloS One* 7 (6): e38011.

"CCPi Tomographic Imaging." 2022. 2022. http://www.ccpi.ac.uk/.

"Cell Tracking Benchmark." 2022. 2022. http://celltrackingchallenge.net/latest-ctb-results/.

Chamier, Lucas von, Romain F. Laine, Johanna Jukkala, Christoph Spahn, Daniel Krentzel, Elias Nehme, Martina Lerche, et al. 2021. "Democratising Deep Learning for Microscopy with ZeroCostDL4Mic." *Nature Communications* 12 (1): 2276.

Chaumont, Fabrice de, Stéphane Dallongeville, Nicolas Chenouard, Nicolas Hervé, Sorin Pop, Thomas Provoost, Vannary Meas-Yedid, et al. 2012. "Icy: An Open Bioimage Informatics Platform for Extended Reproducible Research." *Nature Methods* 9 (7): 690–96.

Chiaruttini, Nicolas, Olivier Burri, Peter Haub, Romain Guiet, Jessica Sordet-Dessimoz, and Arne Seitz. 2022. "An Open-Source Whole Slide Image Registration Workflow at Cellular Precision Using Fiji, QuPath and Elastix." *Frontiers in Computer Science* 3 (January). https://doi.org/10.3389/fcomp.2021.780026.

Culley, Siân, David Albrecht, Caron Jacobs, Pedro Matos Pereira, Christophe Leterrier, Jason Mercer, and Ricardo Henriques. 2018. "Quantitative Mapping and Minimization of Super-Resolution Optical Imaging Artifacts." *Nature Methods* 15 (4): 263–66.

"Discourse - Civilized Discussion." 2022. Discourse - Civilized Discussion. 2022. https://www.discourse.org/.

Domander, Richard, Alessandro A. Felder, and Michael Doube. 2021. "BoneJ2 - Refactoring Established Research Software." *Wellcome Open Research* 6 (February): 37.

Dzmitry Malyshau, Kai Ninomiya, Brandon Jones (Editors). 2022. "WebGPU." 2022. https://www.w3.org/TR/webgpu/.

Edelstein, Arthur, Nenad Amodaj, Karl Hoover, Ron Vale, and Nico Stuurman. 2010. "Computer Control of Microscopes Using μManager." *Current Protocols in Molecular Biology / Edited by Frederick M. Ausubel ... [et Al.]* Chapter 14 (October): Unit14.20.

Elixir community. 2022. "Bio.tools." Bio.tools. 2022. https://bio.tools/.

Fedorov, Andriy, Reinhard Beichel, Jayashree Kalpathy-Cramer, Julien Finet, Jean-Christophe Fillion-Robin, Sonia Pujol, Christian Bauer, et al. 2012. "3D Slicer as an Image Computing Platform for the Quantitative Imaging Network." *Magnetic Resonance Imaging* 30 (9): 1323–41.

Gibbs, Holly C., Sakina M. Mota, Nathan A. Hart, Sun Won Min, Alex O. Vernino, Anna L. Pritchard, Anindito Sen, et al. 2021. "Navigating the Light-Sheet Image Analysis Software Landscape: Concepts for Driving Cohesion From Data Acquisition to Analysis." *Frontiers in*

*Cell and Developmental Biology* 9 (November): 739079.

Gómez-de-Mariscal, Estibaliz, Carlos García-López-de-Haro, Wei Ouyang, Laurène Donati, Emma Lundberg, Michael Unser, Arrate Muñoz-Barrutia, and Daniel Sage. 2021. "DeepImageJ: A User-Friendly Environment to Run Deep Learning Models in ImageJ." *Nature Methods* 18 (10): 1192–95.

Gürsoy, Doǧa, Francesco De Carlo, Xianghui Xiao, and Chris Jacobsen. 2014. "TomoPy: A Framework for the Analysis of Synchrotron Tomographic Data." *Journal of Synchrotron Radiation* 21 (Pt 5): 1188–93.

Gustafsson, Nils, Siân Culley, George Ashdown, Dylan M. Owen, Pedro Matos Pereira, and Ricardo Henriques. 2016. "Fast Live-Cell Conventional Fluorophore Nanoscopy with ImageJ through Super-Resolution Radial Fluctuations." *Nature Communications* 7 (August): 12471.

Haase, Robert. 2021. "Image Processing Filters for Grids of Cells Analogous to Filters Processing Grids of Pixels." *Frontiers in Computer Science* 3 (November). https://doi.org/10.3389/fcomp.2021.774396.

Haase, Robert, Akanksha Jain, Stéphane Rigaud, Daniela Vorkel, Pradeep Rajasekhar, Theresa Suckert, Talley J. Lambert, et al. 2020. "Interactive Design of GPU-Accelerated Image Data Flow Graphs and Cross-Platform Deployment Using Multi-Lingual Code Generation." *bioRxiv*. https://doi.org/10.1101/2020.11.19.386565.

Haase, Robert, Loic A. Royer, Peter Steinbach, Deborah Schmidt, Alexandr Dibrov, Uwe Schmidt, Martin Weigert, et al. 2020. "CLIJ: GPU-Accelerated Image Processing for Everyone." *Nature Methods* 17 (1): 5–6.

Harris, Charles R., K. Jarrod Millman, Stéfan J. van der Walt, Ralf Gommers, Pauli Virtanen, David Cournapeau, Eric Wieser, et al. 2020. "Array Programming with NumPy." *Nature* 585 (7825): 357–62.

Helmstaedter, Moritz, Kevin L. Briggman, and Winfried Denk. 2011. "High-Accuracy Neurite Reconstruction for High-Throughput Neuroanatomy." *Nature Neuroscience* 14 (8): 1081–88.

Hörl, David, Fabio Rojas Rusak, Friedrich Preusser, Paul Tillberg, Nadine Randel, Raghav K. Chhetri, Albert Cardona, et al. 2019. "BigStitcher: Reconstructing High-Resolution Image Datasets of Cleared and Expanded Samples." *Nature Methods* 16 (9): 870–74.

Hunter, John D. 2007. "Matplotlib: A 2D Graphics Environment." *Computing in Science & Engineering* 9 (3): 90–95.

Hu, Yuzhi, Ajay Limaye, and Jing Lu. 2020. "Three-Dimensional Segmentation of Computed Tomography Data Using : New Tools and Developments." *Royal Society Open Science* 7 (12): 201033.

Jones, Thouis R., In Han Kang, Douglas B. Wheeler, Robert A. Lindquist, Adam Papallo, David M. Sabatini, Polina Golland, and Anne E. Carpenter. 2008. "CellProfiler Analyst: Data Exploration and Analysis Software for Complex Image-Based Screens." *BMC Bioinformatics* 9 (November): 482.

Klein, Stefan, Marius Staring, Keelin Murphy, Max A. Viergever, and Josien P. W. Pluim. 2010. "Elastix: A Toolbox for Intensity-Based Medical Image Registration." *IEEE Transactions on Medical Imaging* 29 (1): 196–205.

Kluyver, Thomas, Benjamin Ragan-Kelley, Fernando Pérez, Brian Granger, Matthias Bussonnier, Jonathan Frederic, Kyle Kelley, et al. 2016. "Jupyter Notebooks – a Publishing Format for Reproducible Computational Workflows." In *Positioning and Power in Academic Publishing: Players, Agents and Agendas*, 87–90. IOS Press.

Kremer, J. R., D. N. Mastronarde, and J. R. McIntosh. 1996. "Computer Visualization of Three-Dimensional Image Data Using IMOD." *Journal of Structural Biology* 116 (1): 71–76.

Krull, Alexander, Tim-Oliver Buchholz, and Florian Jug. 2018. "Noise2Void - Learning Denoising from Single Noisy Images." http://arxiv.org/abs/1811.10980.

Laine, Romain F., Kalina L. Tosheva, Nils Gustafsson, Robert D. M. Gray, Pedro Almada, David Albrecht, Gabriel T. Risa, et al. 2019. "NanoJ: A High-Performance Open-Source Super-Resolution Microscopy Toolbox." *Journal of Physics D: Applied Physics* 52 (16): 163001.

Legland, David, Ignacio Arganda-Carreras, and Philippe Andrey. 2016. "MorphoLibJ: Integrated Library and Plugins for Mathematical Morphology with ImageJ." *Bioinformatics* 32 (22): 3532–34.

Levet, Florian, Anne E. Carpenter, Kevin W. Eliceiri, Anna Kreshuk, Peter Bankhead, and Robert Haase. 2021. "Developing Open-Source Software for Bioimage Analysis: Opportunities and Challenges." *F1000Research* 10 (April): 302.

Levet, Florian, Eric Hosy, Adel Kechkar, Corey Butler, Anne Beghin, Daniel Choquet, and Jean-Baptiste Sibarita. 2015. "SR-Tesseler: A Method to Segment and Quantify Localization-Based Super-Resolution Microscopy Data." *Nature Methods* 12 (11): 1065–71.

Levin, Barnaby D. A., Yi Jiang, Elliot Padgett, Shawn Waldon, Cory Quammen, Chris Harris, Utkarsh Ayachit, et al. 2018. "Tutorial on the Visualization of Volumetric Data Using Tomviz." *Microscopy Today* 26 (1): 12–17.

Limaye, Ajay. 2012. "Drishti: A Volume Exploration and Presentation Tool." In *Developments in X-Ray Tomography VIII*, edited by Stuart R. Stock. SPIE. https://doi.org/10.1117/12.935640.

Linkert, Melissa, Curtis T. Rueden, Chris Allan, Jean-Marie Burel, Will Moore, Andrew Patterson, Brian Loranger, et al. 2010. "Metadata Matters: Access to Image Data in the Real World." *The Journal of Cell Biology* 189 (5): 777–82.

Lobet, Guillaume, Xavier Draye, and Claire Périlleux. 2013. "An Online Database for Plant Image Analysis Software Tools." *Plant Methods* 9 (1): 38.

"Macro Language." 2022. 2022. https://imagej.nih.gov/ij/developer/macro/macros.html.

Marée, Raphaël, Loïc Rollus, Benjamin Stévens, Renaud Hoyoux, Gilles Louppe, Rémy Vandaele, Jean-Michel Begon, Philipp Kainz, Pierre Geurts, and Louis Wehenkel. 2016. "Collaborative Analysis of Multi-Gigapixel Imaging Data Using Cytomine." *Bioinformatics* 32 (9): 1395–1401.

Marsh, Richard J., Ishan Costello, Mark-Alexander Gorey, Donghan Ma, Fang Huang, Mathias Gautel, Maddy Parsons, and Susan Cox. 2021. "Sub-Diffraction Error Mapping for Localisation Microscopy Images." *Nature Communications* 12 (1): 5611.

Mastronarde, David N. 2005. "Automated Electron Microscope Tomography Using Robust Prediction of Specimen Movements." *Journal of Structural Biology* 152 (1): 36–51.

Mathis, Alexander, Pranav Mamidanna, Kevin M. Cury, Taiga Abe, Venkatesh N. Murthy, Mackenzie Weygandt Mathis, and Matthias Bethge. 2018. "DeepLabCut: Markerless Pose Estimation of User-Defined Body Parts with Deep Learning." *Nature Neuroscience* 21 (9): 1281–89.

McQuin, Claire, Allen Goodman, Vasiliy Chernyshev, Lee Kamentsky, Beth A. Cimini, Kyle W. Karhohs, Minh Doan, et al. 2018. "CellProfiler 3.0: Next-Generation Image Processing for Biology." *PLoS Biology* 16 (7): e2005970.

Meijering, E., M. Jacob, J-C F. Sarria, P. Steiner, H. Hirling, and M. Unser. 2004. "Design and Validation of a Tool for Neurite Tracing and Analysis in Fluorescence Microscopy Images." *Cytometry. Part A: The Journal of the International Society for Analytical Cytology* 58 (2): 167–76.

Miles, Alistair, John Kirkham, Martin Durant, James Bourbeau, Tarik Onalan, Joe Hamman, Zain Patel, et al. 2020. "Zarr-Developers/zarr-Python: v2.4.0," January. https://doi.org/10.5281/zenodo.3773450.

Miura, Kota, and Simon F. Nørrelykke. 2021. "Reproducible Image Handling and Analysis." *The EMBO Journal* 40 (3): e105889.

Miura, Kota, and Natasa Sladoje, eds. 2019. *Bioimage Data Analysis Workflows*. 1st ed. Learning Materials in Biosciences. Cham, Switzerland: Springer Nature.

Moore, Josh, Chris Allan, Sébastien Besson, Jean-Marie Burel, Erin Diel, David Gault, Kevin Kozlowski, et al. 2021. "OME-NGFF: A next-Generation File Format for Expanding Bioimaging Data-Access Strategies." *Nature Methods* 18 (12): 1496–98.

Müller, Marcel, Viola Mönkemöller, Simon Hennig, Wolfgang Hübner, and Thomas Huser. 2016. "Open-Source Image Reconstruction of Super-Resolution Structured Illumination Microscopy Data in ImageJ." *Nature Communications* 7 (March): 10980.
NEUBIAS. 2020. "NEUBIAS Academy Youtube Channel." 2020. https://www.youtube.com/neubias.
Ollion, Jean, Julien Cochennec, François Loll, Christophe Escudé, and Thomas Boudier. 2013. "TANGO: A Generic Tool for High-Throughput 3D Image Analysis for Studying Nuclear Organization." *Bioinformatics* 29 (14): 1840–41.
Ouyang, Wei, Florian Mueller, Martin Hjelmare, Emma Lundberg, and Christophe Zimmer. 2019. "ImJoy: An Open-Source Computational Platform for the Deep Learning Era." *Nature Methods* 16 (12): 1199–1200.
Ovesný, Martin, Pavel Křížek, Josef Borkovec, Zdeněk Svindrych, and Guy M. Hagen. 2014. "ThunderSTORM: A Comprehensive ImageJ Plug-in for PALM and STORM Data Analysis and Super-Resolution Imaging." *Bioinformatics* 30 (16): 2389–90.
Palla, Giovanni, Hannah Spitzer, Michal Klein, David Sebastian Fischer, Anna Christina Schaar, Louis Benedikt Kuemmerle, Sergei Rybakov, et al. 2021. "Squidpy: A Scalable Framework for Spatial Single Cell Analysis." *bioRxiv*. bioRxiv. https://doi.org/10.1101/2021.02.19.431994.
Paszke, Adam, Sam Gross, Francisco Massa, Adam Lerer, James Bradbury, Gregory Chanan, Trevor Killeen, et al. 2019. "PyTorch: An Imperative Style, High-Performance Deep Learning Library." http://arxiv.org/abs/1912.01703.
Pau, Grégoire, Florian Fuchs, Oleg Sklyar, Michael Boutros, and Wolfgang Huber. 2010. "EBImage--an R Package for Image Processing with Applications to Cellular Phenotypes." *Bioinformatics* 26 (7): 979–81.
Pedregosa, Fabian, Gaël Varoquaux, Alexandre Gramfort, Vincent Michel, Bertrand Thirion, Olivier Grisel, Mathieu Blondel, et al. 2011. "Scikit-Learn: Machine Learning in Python." *Journal of Machine Learning Research: JMLR* 12 (85): 2825–30.
Perkel, Jeffrey M. 2021. "Python Power-up: New Image Tool Visualizes Complex Data." *Nature* 600 (7888): 347–48.
Pielawski, Nicolas, Axel Andersson, Christophe Avenel, Andrea Behanova, Eduard Chelebian, Anna Klemm, Fredrik Nysjö, Leslie Solorzano, and Carolina Wählby. 2022. "TissUUmaps 3: Interactive Visualization and Quality Assessment of Large-Scale Spatial Omics Data." https://doi.org/10.1101/2022.01.28.478131.
Pietzsch, Tobias, Stephan Preibisch, Pavel Tomancák, and Stephan Saalfeld. 2012. "ImgLib2--Generic Image Processing in Java." *Bioinformatics* 28 (22): 3009–11.
Pietzsch, Tobias, Stephan Saalfeld, Stephan Preibisch, and Pavel Tomancak. 2015. "BigDataViewer: Visualization and Processing for Large Image Data Sets." *Nature Methods* 12 (6): 481–83.
Pinkard, Henry, Nico Stuurman, Ivan E. Ivanov, Nicholas M. Anthony, Wei Ouyang, Bin Li, Bin Yang, et al. 2021. "Pycro-Manager: Open-Source Software for Customized and Reproducible Microscope Control." *Nature Methods* 18 (3): 226–28.
Püspöki, Zsuzsanna, Martin Storath, Daniel Sage, and Michael Unser. 2016. "Transforms and Operators for Directional Bioimage Analysis: A Survey." *Advances in Anatomy, Embryology, and Cell Biology* 219: 69–93.
"Python Package Index Download Statistics." 2021. 2021. https://pypistats.org/top.
Ries, Jonas. 2020. "SMAP: A Modular Super-Resolution Microscopy Analysis Platform for SMLM Data." *Nature Methods* 17 (9): 870–72.
Rocklin, Matthew. 2015. "Dask: Parallel Computation with Blocked Algorithms and Task Scheduling." In *Proceedings of the 14th Python in Science Conference*. SciPy. https://doi.org/10.25080/majora-7b98e3ed-013.
Rubens, Ulysse, Renaud Hoyoux, Laurent Vanosmael, Mehdy Ouras, Maxime Tasset,

Christopher Hamilton, Rémi Longuespée, and Raphaël Marée. 2019. "Cytomine: Toward an Open and Collaborative Software Platform for Digital Pathology Bridged to Molecular Investigations." *Proteomics. Clinical Applications* 13 (1): e1800057.

Rubens, Ulysse, Romain Mormont, Lassi Paavolainen, Volker Bäcker, Benjamin Pavie, Leandro A. Scholz, Gino Michiels, et al. 2020. "BIAFLOWS: A Collaborative Framework to Reproducibly Deploy and Benchmark Bioimage Analysis Workflows." *Patterns (New York, N.Y.)* 1 (3): 100040.

Rueden, Curtis T., Jeanelle Ackerman, Ellen T. Arena, Jan Eglinger, Beth A. Cimini, Allen Goodman, Anne E. Carpenter, and Kevin W. Eliceiri. 2019. "Scientific Community Image Forum: A Discussion Forum for Scientific Image Software." *PLoS Biology* 17 (6): e3000340.

Rueden, Curtis T., Johannes Schindelin, Mark C. Hiner, Barry E. DeZonia, Alison E. Walter, Ellen T. Arena, and Kevin W. Eliceiri. 2017. "ImageJ2: ImageJ for the next Generation of Scientific Image Data." *BMC Bioinformatics* 18 (1): 529.

Ryosuke Okuta, Yuya Unno, Daisuke Nishino, Shohei Hido, Crissman Loomis. 2017. "CuPy: A NumPy-Compatible Library for NVIDIA GPU Calculations." Proceedings of Workshop on Machine Learning Systems (LearningSys) in The Thirty-First Annual Conference on Neural Information Processing Systems (NIPS). 2017. http://learningsys.org/nips17/assets/papers/paper_16.pdf.

Saalfeld, Stephan, Albert Cardona, Volker Hartenstein, and Pavel Tomancak. 2009. "CATMAID: Collaborative Annotation Toolkit for Massive Amounts of Image Data." *Bioinformatics* 25 (15): 1984–86.

Sage, Daniel, Lauréne Donati, Ferréol Soulez, Denis Fortun, Guillaume Schmit, Arne Seitz, Romain Guiet, Cédric Vonesch, and Michael Unser. 2017. "DeconvolutionLab2: An Open-Source Software for Deconvolution Microscopy." *Methods* 115 (February): 28–41.

Sage, Daniel, Thanh-An Pham, Hazen Babcock, Tomas Lukes, Thomas Pengo, Jerry Chao, Ramraj Velmurugan, et al. 2019. "Super-Resolution Fight Club: Assessment of 2D and 3D Single-Molecule Localization Microscopy Software." *Nature Methods* 16 (5): 387–95.

Scheres, Sjors H. W. 2012. "RELION: Implementation of a Bayesian Approach to Cryo-EM Structure Determination." *Journal of Structural Biology* 180 (3): 519–30.

Schindelin, Johannes, Ignacio Arganda-Carreras, Erwin Frise, Verena Kaynig, Mark Longair, Tobias Pietzsch, Stephan Preibisch, et al. 2012. "Fiji: An Open-Source Platform for Biological-Image Analysis." *Nature Methods* 9 (7): 676–82.

Schmidt, Uwe, Martin Weigert, Coleman Broaddus, and Gene Myers. 2018. "Cell Detection with Star-Convex Polygons." In *Medical Image Computing and Computer Assisted Intervention – MICCAI 2018*, 265–73. Lecture Notes in Computer Science. Cham: Springer International Publishing.

Schneider, Caroline A., Wayne S. Rasband, and Kevin W. Eliceiri. 2012. "NIH Image to ImageJ: 25 Years of Image Analysis." *Nature Methods* 9 (7): 671–75.

"SciJava." 2022. 2022. https://scijava.org/.

Sofroniew, Nicholas, Talley Lambert, Kira Evans, Juan Nunez-Iglesias, Grzegorz Bokota, Gonzalo Peña-Castellanos, Philip Winston, et al. 2021. *Napari/napari: 0.4.12rc2*. Zenodo. https://doi.org/10.5281/ZENODO.3555620.

"Software Tools For Molecular Microscopy." 2006. 2006. https://en.wikibooks.org/wiki/Software_Tools_For_Molecular_Microscopy.

Solorzano, Leslie, Gabriele Partel, and Carolina Wählby. 2020. "TissUUmaps: Interactive Visualization of Large-Scale Spatial Gene Expression and Tissue Morphology Data." *Bioinformatics* 36 (15): 4363–65.

"Stack Overflow Developer Survey 2020." 2020. 2020. https://insights.stackoverflow.com/survey/2020/?utm_source=social-share&utm_medium=social&utm_campaign=dev-survey-2020.

"Stack Overflow Developer Survey 2021." 2021. Stack Overflow. 2021.

https://insights.stackoverflow.com/survey/2021/?utm_source=social-share&utm_medium=social&utm_campaign=dev-survey-2021.

Stephan Saalfeld, Igor Pisarev, Philipp Hanslovsky, John Bogovic, Andrew Champion, Curtis Rueden, John Kirkham. 2017. "N5: Not HDF5." 2017. https://github.com/saalfeldlab/n5.

Stoltzfus, Caleb R., Jakub Filipek, Benjamin H. Gern, Brandy E. Olin, Joseph M. Leal, Yajun Wu, Miranda R. Lyons-Cohen, et al. 2020. "CytoMAP: A Spatial Analysis Toolbox Reveals Features of Myeloid Cell Organization in Lymphoid Tissues." *Cell Reports* 31 (3): 107523.

Stringer, Carsen, Tim Wang, Michalis Michaelos, and Marius Pachitariu. 2021. "Cellpose: A Generalist Algorithm for Cellular Segmentation." *Nature Methods* 18 (1): 100–106.

Tang, Guang, Liwei Peng, Philip R. Baldwin, Deepinder S. Mann, Wen Jiang, Ian Rees, and Steven J. Ludtke. 2007. "EMAN2: An Extensible Image Processing Suite for Electron Microscopy." *Journal of Structural Biology* 157 (1): 38–46.

Tensorflow-Developers. 2021. *TensorFlow*. Zenodo. https://doi.org/10.5281/ZENODO.4724125.

"The Apache Groovy Programming Language." 2022. 2022. https://groovy-lang.org/.

"The R Project for Statistical Computing." 2022. 2022. https://www.R-project.org/.

———. n.d. Accessed January 23, 2022. https://www.R-project.org/.

Thévenaz, P., U. E. Ruttimann, and M. Unser. 1998. "A Pyramid Approach to Subpixel Registration Based on Intensity." *IEEE Transactions on Image Processing: A Publication of the IEEE Signal Processing Society* 7 (1): 27–41.

Tinevez, Jean-Yves, Nick Perry, Johannes Schindelin, Genevieve M. Hoopes, Gregory D. Reynolds, Emmanuel Laplantine, Sebastian Y. Bednarek, Spencer L. Shorte, and Kevin W. Eliceiri. 2017. "TrackMate: An Open and Extensible Platform for Single-Particle Tracking." *Methods* 115 (February): 80–90.

Tischer, Christian, Ashis Ravindran, Sabine Reither, Nicolas Chiaruttini, Rainer Pepperkok, and Nils Norlin. 2021. "BigDataProcessor2: A Free and Open-Source Fiji Plugin for Inspection and Processing of TB Sized Image Data." *Bioinformatics*, February. https://doi.org/10.1093/bioinformatics/btab106.

Ulman, Vladimír, Martin Maška, Klas E. G. Magnusson, Olaf Ronneberger, Carsten Haubold, Nathalie Harder, Pavel Matula, et al. 2017. "An Objective Comparison of Cell-Tracking Algorithms." *Nature Methods* 14 (12): 1141–52.

Vergara, Hernando M., Constantin Pape, Kimberly I. Meechan, Valentyna Zinchenko, Christel Genoud, Adrian A. Wanner, Kevin Nzumbi Mutemi, et al. 2021. "Whole-Body Integration of Gene Expression and Single-Cell Morphology." *Cell* 184 (18): 4819–37.e22.

Virtanen, Pauli, Ralf Gommers, Travis E. Oliphant, Matt Haberland, Tyler Reddy, David Cournapeau, Evgeni Burovski, et al. 2020. "SciPy 1.0: Fundamental Algorithms for Scientific Computing in Python." *Nature Methods* 17 (3): 261–72.

Walt, Stéfan van der, Johannes L. Schönberger, Juan Nunez-Iglesias, François Boulogne, Joshua D. Warner, Neil Yager, Emmanuelle Gouillart, Tony Yu, and scikit-image contributors. 2014. "Scikit-Image: Image Processing in Python." *PeerJ* 2 (June): e453.

Wei, Donglai, Zudi Lin, Daniel Franco-Barranco, Nils Wendt, Xingyu Liu, Wenjie Yin, Xin Huang, et al. 2020. "MitoEM Dataset: Large-Scale 3D Mitochondria Instance Segmentation from EM Images." *Medical Image Computing and Computer-Assisted Intervention: MICCAI ... International Conference on Medical Image Computing and Computer-Assisted Intervention* 12265 (October): 66–76.

Weigert, Martin, Uwe Schmidt, Tobias Boothe, Andreas Müller, Alexandr Dibrov, Akanksha Jain, Benjamin Wilhelm, et al. 2018. "Content-Aware Image Restoration: Pushing the Limits of Fluorescence Microscopy." *Nature Methods* 15 (12): 1090–97.

Weigert, Martin, Uwe Schmidt, Robert Haase, Ko Sugawara, and Gene Myers. 2020. "Star-Convex Polyhedra for 3D Object Detection and Segmentation in Microscopy." In *2020 IEEE Winter Conference on Applications of Computer Vision (WACV)*. IEEE. https://doi.org/10.1109/wacv45572.2020.9093435.

Witten, Ian H., Eibe Frank, Mark A. Hall, and Christopher J. Pal. 2016. *Data Mining: Practical Machine Learning Tools and Techniques*. Morgan Kaufmann.

Yoo, Terry S., Michael J. Ackerman, William E. Lorensen, Will Schroeder, Vikram Chalana, Stephen Aylward, Dimitris Metaxas, and Ross Whitaker. 2002. "Engineering and Algorithm Design for an Image Processing Api: A Technical Report on ITK--the Insight Toolkit." *Studies in Health Technology and Informatics* 85: 586–92.

Yushkevich, Paul A., Joseph Piven, Heather Cody Hazlett, Rachel Gimpel Smith, Sean Ho, James C. Gee, and Guido Gerig. 2006. "User-Guided 3D Active Contour Segmentation of Anatomical Structures: Significantly Improved Efficiency and Reliability." *NeuroImage* 31 (3): 1116–28.